# Carbon Film in Radio Frequency Surface Plasma Source with Cesiation

Vadim Dudnikov[1a)], B. Han[2], S. Murray[2], T. Pennisi[2],
C. Stinson[2], M. Stockli[2], R. Welton[2], A. Dudnikov[3]

[1]Muons, Inc, 552 N. Batavia Ave. Batavia, IL,60510, USA,
[2] ORNL, Oak Ridge, TN 37830, USA, [3]Novosibirsk, Russia, 630090

[a)] Corresponding author: Vadim@muonsinc.com

It is assumed that persistent cesiation in the SNS RF SPS is related to deposition of carbon film on the collar converter. The work function dependence for graphite with alkali deposition has no minimum typical for metals and semiconductors and the final work function is higher. For this reason, the probability of H- secondary emission from cesiated metal and semiconductors can be higher than from cesiated carbon films but the carbon film maintains cesiation longer and can operate with low cesium consumption.

## INTRODUCTION

A physical basis for high intense negative ion beam production now is a "cesiation effect".
The cesiation effect [[1],[23],[4],[5]], a significant enhancement of negative ion emission from gas discharges with decrease of co-extracted electron current below negative ion current after adding into discharge a small admixture of cesium (or other elements with low ionization potential) is connected with decreases the surface work function (WF) from 4–5 eV to ~2-1.6 eV by cesium adsorption which enhances secondary emission of negative ions caused by the interaction of the plasma with the electrode surface and thereby enhances surface plasma generation (SPG) of negative ions. Ion sources based on this process have been named surface plasma sources (SPSs). With decrease of the WF is increased a probability for escaping as negative ions for desorbed, reflected and sputtered particles. For efficient negative ion generation it is necessary to keep the minimal surface WF contacting with gas discharge plasma. Normally, H− ion beams from cesiated SPS decay unless there is a continuous feed of Cs (from 0.5 mg/day in DESY magnetron SPS to ~1 g/day in Los Alamos converter SPS; more common ~1mg/hour in SPS for accelerators and several gram/run for large volume SPS for fusion neutral beam injectors. Some time impurities with low ionization potential could be extracted by discharge plasma from discharge chamber elements ("cesiation" without cesium).
Fortunately, in RF SPS developed for SNS were found conditions for long time operation with adding a small amount of cesium during beginning operation [[6]].

Long time question: why this cesiation is persistent without the continuous feed Cs?

> One possibility: "Apparently, the plasma conditioning scrubs the metal surface atomically clean, replacing the covalent bonds with surface sorbates with ionic bonds with the metal surface. The persistence of the SNS beam suggests that most Cs sticks well to the converter and therefore is not sputtered by the ultra high purity hydrogen plasma. However, earlier experience suggests that the Cs does not stick well on surface contaminants and is lost within a few hours". [[7],[8]]

Other possibility:
It is very probably that a long lifetime of cesiation in SNS RF surface plasma sources (SPS) can be connected with deposition into the emitter/converter cone and to the discharge chamber some specific carbon films [[9]].
The work function dependence for graphite with alkali deposition has no minimum typical for metals and semiconductors and a final work function is higher. By this reason the probability of H- secondary emission from cesiated metal and semiconductors can be higher than from cesiated carbon films but lest can better keep the cesiation and can operate with low cesium consumption.

It is known that a two-dimensional graphite films and films of pyrolitic graphite can adsorb and trap alkali atoms with very high probability (sticking coefficient ~1) up to very high temperature and can be desorbed by heating up to high temperature, much higher than for evaporation from a metal surface (up to 2000K). A dynamic of carbon film formation and alkali atoms trapping is well investigated in high vacuum conditions, but in SPS these processes are complicated by high gas density and by gas discharge plasma. Investigation of these processes in condition of real SPS can be important for improve of SPS performances.

Intercalation of different elements in two-dimensional graphite films on metals was observed in 1981 [10] and described it in detail in reviews [11,12,13,14]. It was found that 2DGFs are formed both on the surface of many metals that give no bulk carbides (Pt, Ni, Re, Ir, Ru, Rh, and Pd) and on bulk carbides of *d*-metals (Mo, Ta, Nb, Ti, Zr, and Hf) without commensurability with the substrate, which exists only for the Ni (111) face. These results are explained by the extremely high adsorption and chemical inertness of the graphite layer, which is bound to the substrate by only weak van der Waals forces and behaves like a two-dimensional crystal. However, the carbon atoms inside the layer are bound very strongly and the film itself has the same atomic structure as a layer in a graphite single crystal.

When intercalated separately, atoms with low ionization potentials (Cs, K, Na,…) intercalate into a 2DGF in a virtually similar way: at room temperature the flux of adsorbed atoms is divided into two approximately equal parts. One-half of the atoms penetrates under the layer and is accumulated there in the intercalated state (thermo-desorption γ-phase), and the second half remains adsorbed on the outer film surface (α-phase). The limiting concentrations of electropositive atoms in both phases are on the order of one monolayer, ~ $4 \cdot 10^{14}$ cm$^{-2}$, for both Cs and K. The heating of such a system leads to sequential thermo-desorption of atoms: atoms of the α-phase fully leave the surface at $T$ ~ 700–800 K. Conversely, the intercalated atoms (γ-phase) turn out to be immured under the graphite layer, thus preventing thermo-desorption, and reside under it up to record high temperatures of ~ 2000 K and up. These atoms are desorbed only upon thermal destruction of graphite islands, which proceeds from their edges.

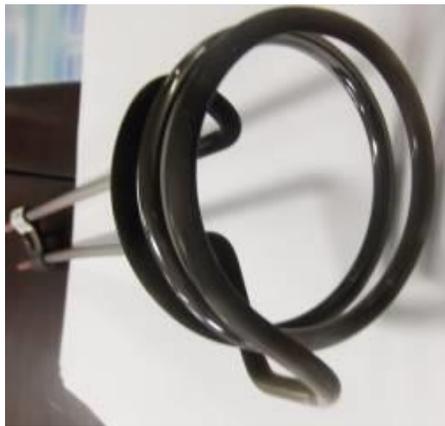
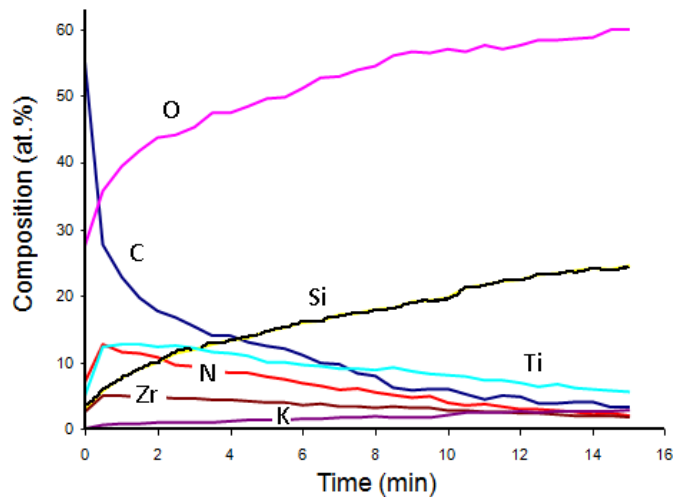

**FIGURE 1**: Internal antenna of baseline SNS RF SPS after long time operation is coated by black film.
**FIGURE 2:** Concentration profiles of elements at the antenna insulation surface (the profile from a dark area on the porcelain). 1 minute of ion sputtering is relating to removing ~20 nm of material.

After long time operation an internal antenna of base line SNS RF H- SPS is coated by black film as shown in Fig. 1. The Figs. 2, show the composition elements profiles from a dark area on the porcelain. The Fig. 2 presents the change in composition as a function of sputter time for all the elements examined (C, Si, O, Ti, Zr, and K). Note that the time scale can be converted to depth scale using a sputter rate of 20 nm/min determined by measuring a standard SiO$_2$ film, which is a reasonable standard for porcelain. Carbon is reduced from ~93 at.% to <30 at.% during the first minute of sputtering and continues to decrease with further sputter, reaching a value of less than 5 at.% by the end of the profile. The maximum K level measured is ~8 at% (which more than enough for enhanced surface plasma generation of H-). In RF discharge the antenna is coated by dense film of carbon (may be diamond like film). It prevents from further sputtering of insulating layer.

Other surfaces of the discharge chamber are coated by carbon to. The converter cone surface was also coated by dark film, possible carbon, as shown in Fig. 3a. Hydrogen saturated diamond film after cesiation can have a negative electron affinity (work function). A formation of such film can be important for efficient formation of secondary negative ions and for a prolonger cesiation lifetime. However, after termination of diamond-like film by hydrogen the cesium adsorption is minimized. May be the plasma presence can change situation.

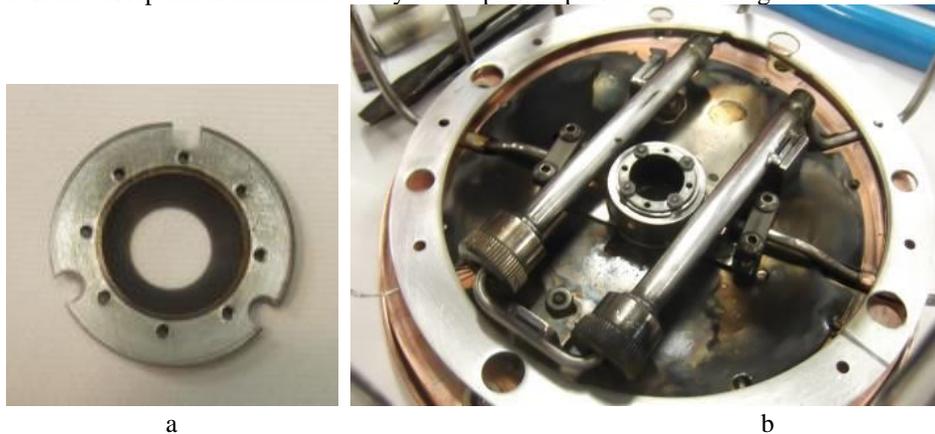

a  b

**FIGURE 3 -a-** Dark film deposition on the converter cone,
b- Dark film deposition on the plasma plate components (surfaces bombarded by intense fluxes of plasma are clean).

Other components of the plasma plate are coated by dark carbon film as shown in Fig. 3b. An optimized deposition of carbon film during conditioning process can be used for preventing the antenna insulating sputtering and for improve the emission properties of converter. Diamondlike film could be deposited to the cold surface with ion assisted deposition. Example of deposition of the transparent, hard carbon film on the St. St. heat shield screen on the surface shielded from the direct plasma flux by collar with magnetic filter is shown in Fig. 4 a. A transparent (blue) film is hard and have resistivity ~10 Ohm between probes separated by 1 cm distance. Black films are more conductive and soft. Similar deposition conditions should have a conical converter collar surface.

The long conditioning of SPS by discharge is necessary for surface cleaning but also for optimal carbon film deposition, helping to keep cesium and supporting high efficiency of negative ion generation. It is useful to provide an analysis of films deposited to the collar at Fig. 3a and at heat shield at Fig. 3b,4a.

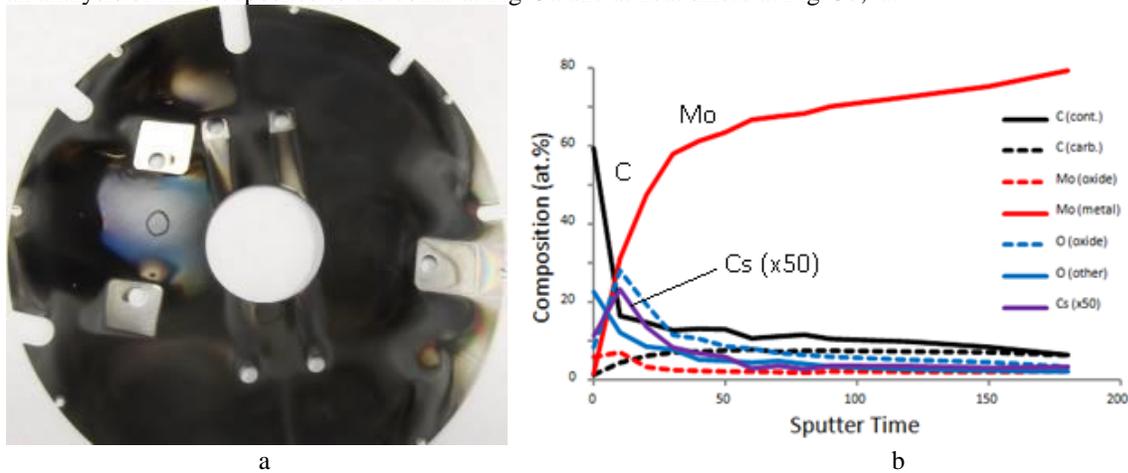

a  b

**FIGURE 4. a-** Deposition of the transparent and black carbon films on the heat shield screen on the surface shielded by collar with magnetic filter.
b- Distributions of elements on the converter cone surface.

It was reviewed a significant number of #3 ion source teardown converters photos from 2010 thru 2012. In every case of this random sampling were found that the Moly converter was darkened. Also were reviewed several that had only run for five days and was somewhat surprised at how dark the converters were. Apparently this carbon buildup happens very early in a source life.

It can be that any specific carbon film is "optimal" for SPS operation. It is useful to provide SEM/EDS, Oge and SIMS analysis of these coatings. Surface analyzing of converter surface is presented in [15].

# PRELIMINARY RESULTS OF FILM DEPOSITION

The converter cone surface was analyzed after long time run in the internal antenna source. The vacuum chamber was filed by Argon gas and ion source was fast moved to the glove box field by argon flow. Ion source was disassembled in Argon and removed collar was transported to the container with Argon, which was pumped for vacuum. A dark deposition is visible on the converter cone surface. The converter cone surface was analyzed by detection of characteristic X-ray photons exited by low energy electrons with surface sputtering by Argon ion beam.

The surface composition detected by characteristic X-Rays is presented in the table below. Distributions of elements on the converter cone surface is presented in Fig. 4 b. Sputter time in minutes. 20 min of sputtering is ~20 nanometers. Carbon concentration is high on the surface as Oxygen in different chemical compounds. Cesium and Sodium are recognizable.

| Surface Composition (at.%) | Mo | O | C | Cs | Na | Cl |
|---|---|---|---|---|---|---|
| As Rec. | 7.6 | 32.8 | 58.6 | 0.3 | 0.5 | 0.3 |
| Sputter Etched | 71.2 | 12.1 | 16.7 | 0.0 | 0.0 | 0.0 |

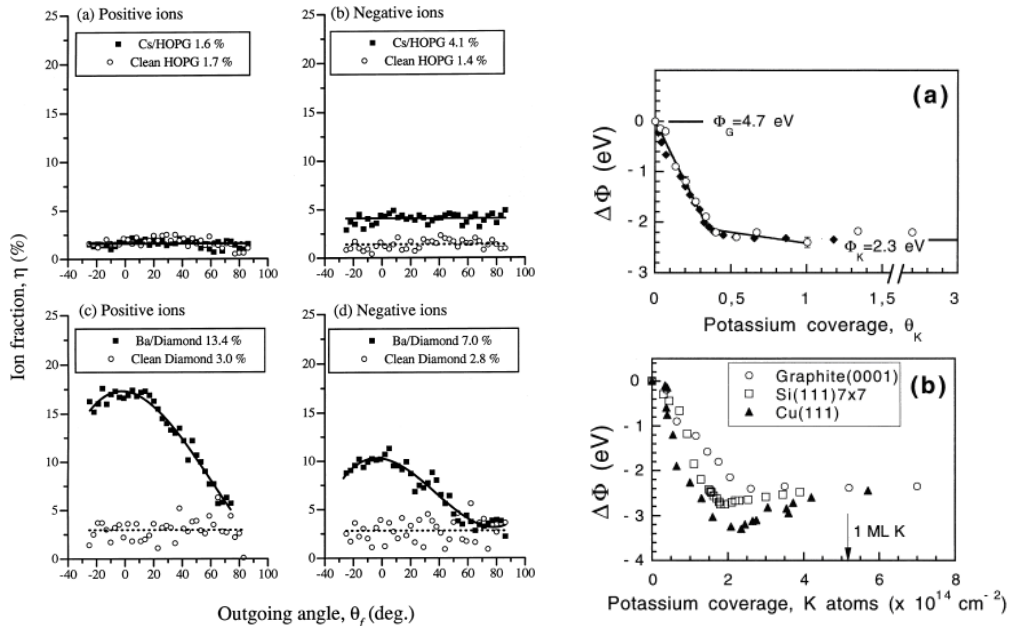

**FIGURE 5**. Influence of Cs deposition on HOPG and Ba deposition to diamond to electron capture of reflected particles ($H_2^+$ with energy ~0.3 keV) [17].

**FIGURE 6.** The work function as a function of potassium coverage measured at $T$=83 K by the RP method (o), and at $T$= 160 K by UPS (diamond) The value of $\Phi_G$ and $\Phi_K$ are taken from Refs. [16] and [17] respectively. The error bars indicate the measurements (most likely due to statistical error of 3 or 4 measurements. (b) A comparison of the work function change as a function of number of K adatoms for graphite, Cu(111), and Si(111) 7×7. The K coverages were obtained by exposures from the same alkali metal source as in estimate the induced dipole moment associated (a), and with surface temperatures of 90 K [18].

Further investigation of surface film deposition is necessary for understanding and reliable reproduce efficient negative ions generation with a low cesium catalyst consumption.

For additional suppression of cesium escaping it is possible to use a laser for resonance excitation of cesium with further ionization by weak RF discharge [19]. Positive Cs ions are trapped in discharge by extraction voltage, extracting the negative ions. Now is not clear, how important is a continious 13.5 MHz discharge in the antenna and converter cone conditioning and deposition and in trapping of cesium ions.

The reflection of protons H+ and negative ion H-from graphite (UOPG) surfaces with deposited Cs and Ba bombarded by $H_2^+$ was measured in [20]. Influential of Cs adsorption on HOPG and Ba adsorption on diamond to electron capture of reflected particles ($H_2^+$ ions with energy ~0.3 keV) is shown in Fig. 5. By deposition of Cs to the HOPG the probability of reflection as H- was increased from $\eta^- =1.4$ to 4.1%. By deposition of Ba to diamond the probability of reflection as H- was increased from $\eta^- =2.8$ to 7%.

It was measured the ion fraction ($\eta$) of the scattered particles leaving the clean and cesiated surfaces to determine if a similar effect occurs for HOPG. Fig. 5a) shows the positive ion fractions H+ (defined in the caption), measured as a function of angle $\theta_f$ for 305 eV $H_2^+$ incident at $\theta_i \sim 70°$ on clean and Cs-dosed HOPG. Fig. 5(b) shows the corresponding negative ion fractions. Within the experimental error, the H+ fraction is unaffected by Cs adsorption. In the case of the scattered H- there is a factor of 3 increase in the ion fraction relative to the clean surface. However, this value is relatively small compared to yields from Cs-covered metal surfaces of comparable work function. The difference in the H- yields may be due to the density of states (DOS) around the HOPG Fermi edge, which is very low in comparison to metals. The electron density is increased by alkali metal adsorption, but remains substantially lower than that of metal surfaces. The charged state of the Cs atoms at the surface may also be a factor. At low coverages, alkali atoms adsorbed on metal surfaces are considered to be partially ionized, setting up a surface dipole layer due to charge transfer to metal surfaces. The characteristic dip that is observed in plots of the surface work function vs. coverage is attributed to the maximum in dipole strength, while further adsorption leads to a metallic alkali overlayer, the removal of the dipole, and a constant work function characteristic of the pure alkali metal. The maximum negative ion fraction from metal surfaces is observed for the minimum work function [12]. In the region of constant work function (i.e., at higher alkali coverages), the H- fraction is ~6%, comparable with the HOPG measurements. Since alkali atoms readily incorporate themselves with the lattice of

HOPG, the charge donation to the surface will be substantially different from that of metals. Coverage dependent work function curves measured for alkali metal adsorption on HOPG do not show the dip observed for metal surfaces [11], indicating that the dipole depolarization model used for metal surfaces may not be applicable to HOPG. In addition, work function measurements are generally a macroscopic property of the surface. However, ion formation may be influenced by work function variations at the local (atomic) level, which may be a substantial difference for Cs atoms adsorbed on HOPG compared to Cs on metal surfaces. As a comparison, Fig. 5 c shows the H+ fractions measured for $H_2^+$ incident clean and Ba dosed polycrystalline diamond. Fig. 5 d shows the corresponding H- fractions. Boron-doped CVD diamond is a semi-conductor with a large band gap (5.5 eV). The ion fractions measured from the clean diamond are somewhat larger than from clean HOPG surface. They are comparable with results obtained by Wurz et al. at more grazing incidence ($\theta_i$ 82°) for scattering of H- from polycrystalline diamond. When Ba is dosed on the diamond surface there is a substantial increase in both the H+ (17% maximum) and the H- (10% maximum) fractions. The high H- fraction from the Ba-dosed polycrystalline diamond sample is consistent with alkali (alkaline earth)-dosed metal surfaces. It can be attributed to the barium being localized on the sample surface, forming a (partial) metal overlayer. Surface localization of a significant amount of the adsorbed Ba is supported by changes to the energy and angular distributions of the scattered hydrogen ions (not shown).

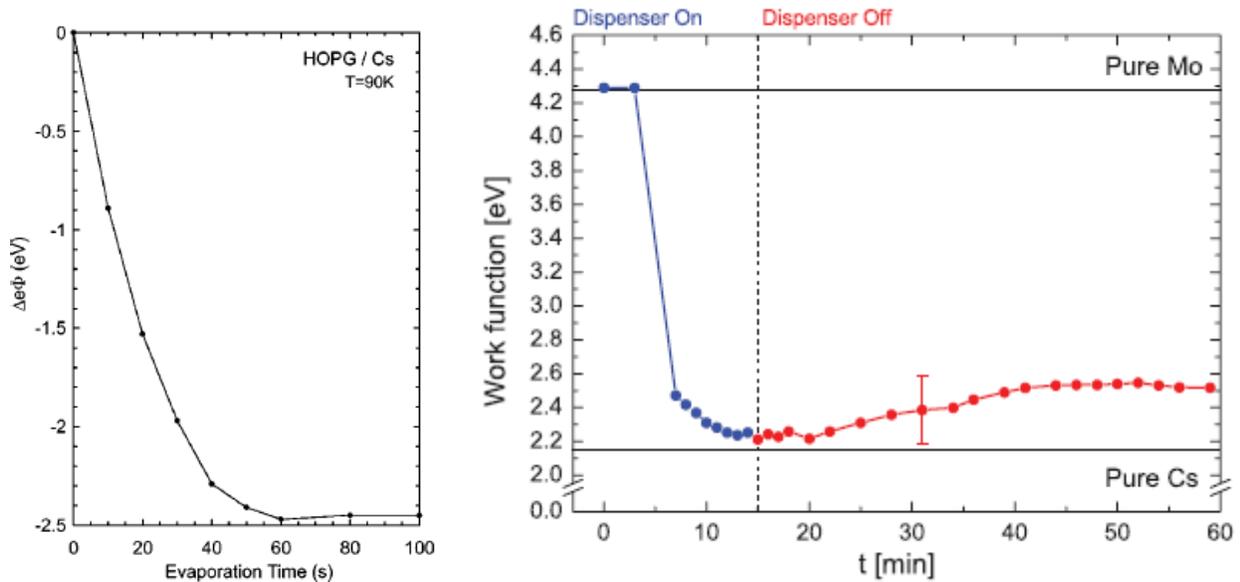

**FIGURE 7. a-** Work function change for graphite upon Cs deposition (1 ML of Cs corresponds to an evaporation time of 350 s) [17].
**b-** Work function time trace of a cesium exposed molybdenum sample under ion source relevant plasma conditions

[23].

For both clean and Cs-dosed HOPG, the ion fractions remain approximately constant across the range of θf measured. This is also true of the clean diamond surface. In contrast, the ion fractions from the Ba-dosed diamond show a strong variation as a function of θf. This is consistent with charge exchange behavior from a metal-like surface, where ions survival is dependent on the outgoing trajectory (i.e. enhanced survival probability along the surface normal). It is remarkable that such an angular dependence is not observed for the (metal-like) HOPG surfaces. This may be due to the anisotropic nature of HOPG influencing the charge transfer process. In addition, the low DOS at the HOPG Fermi edge will have a strong influence the H- survival probability while the hydrogen species are close to the surface. Interestingly, the addition of Cs, which can be expected to increase the metallic nature of the HOPG, is not sufficient to produce an angular dependence similar to that observed for the Ba-dosed diamond sample. In conclusion, the results demonstrate that while the electronic properties of HOPG are modified by Cs adsorption, there are no observable structural changes and no appreciable Cs remaining on the outermost layer. Despite the fact that the work function of the cesiated HOPG is comparable to that of alkali covered metal surfaces, a large increase in H- production is not observed. In contrast, a significant increase in the scattered ion fraction is observed when barium is dosed on a polycrystalline diamond sample. This is attributed to the fact that the Ba remains on the sample surface, acting as a metal overlayer. Dependences of the work function on potassium coverages of graphite, copper and silicon are shown in Fig. 6 from [14]. Alkali metal adsorption on graphite and calculation of a work function variation is presented in Ref. [21].

Work function change for graphite upon Cs deposition (1 ML of Cs corresponds to an evaporation time of 350 s.) is shown in Fig. 7 a from [22].

Very similar dependence of work function was produced during a cesium deposition to the molybdenum processed by RF plasma flux in [23]. Work function time trace of a cesium exposed molybdenum sample under ion source relevant plasma conditions is shown in Fig. 7 b. The work function decreases during a slow deposition of Cs flux until Φ=2.3 eV as in Fig. 7 (a) and increases slowly after removing of cesium source. It is probable that during molybdenum sample plasma processing any carbon films were deposited. It is interesting, that without plasma processing the work function can be decrease by Cs deposition only up to 2.6 eV fare above a WF of thick cesium layer 2.14 eV.

These measurements show that the work function of pure cesium is not achieved, even after several minutes of intense cesium exposure. This indicates the presence of electronegative impurity species resulting in an increase of the work function of the surface. The increase of the work function without cesium exposition indicates a contamination of the given cesium layers by the adsorption and absorption of impurity from the residual gas. This behavior is very similar to the observations reported in Ref. 17 where the contamination of a single monolayer of cesium was observed within a short timescale for even better vacuum conditions of $p = 10^{-6}$ Pa. There was no difference of WF behavior for Cs deposition to the Molybdenum and Molybdenum doped by Lanthanum which can have lower WF without cesium deposition.

For stable and reproducible production of surfaces with lower WF it is necessary to perform specific cleaning and processing of electrode surfaces. For reproducible production of surfaces with low work function, it is necessary to heat them for a long time at high temperatures to remove volume impurities. Fortunately, in SPS from the beginning there was the possibility for high temperature (>1500 K) heating by discharges [24,25]. Discharges with air were used for electrode cleaning of excess cesium and from carbon. Some procedures of low work function production in systems with cesium, similar to SPS are discussed in [26].

In the first versions of compact SPS [19] as a flat magnetron (planotron) negative ions was emitted from cathode bombarded by ions from discharge in crossed field. In discharges without cesium the energy of bombarding ions was ~600 V and cathode was heated up to >1000 C. This processing should be enough for production of a clean metallic surface. After cesiation the discharge voltage decreases up to ~100 V with high current density ~ 100A/cm$^2$. These conditions were enough for sputtering of impurity, including cesium. However, cesium was returned to the cathode as positive ions. With high level of cesium ionization the cesium loss was suppressed by extraction voltage (but cesium can escape from discharge chamber as neutral atoms after discharge). One possibility to suppress this loss is using of a resonance laser for excitation and ionization of cesium between discharge pulses [19]. For compensation of this cesium loss was used a slow injection of cesium by heating of cesium chromate mixture by discharge or by special heater. For vacuum pumping were used diffusion pumps with an oil mechanical pump. A high efficiency of H- production was not suppressed by these carbohydrate contaminations.

SPS parts cooled in vacuum were very clean and were not corrode during very long time.

It is possible that hot surfaces were deposited by thin protecting carbon films during cooling in vacuum contaminated by vacuum oil vapors.

After transition to more cleaner vacuum with Ti getter pumping and with turbovacs was no visible change of this compact SPS operation.

In other SPS with "anode production" of negative ions near the emission aperture (as in Penning discharge and large volume SPS) [3] was no such strong surface cleaning (however it can be heated up to high temperature) and deposition of impurities films become more important. In these sources was more important a long time "conditioning" by discharge. H- emission enhancing by long time conditioning of plasma electrode with analysis of the deposited impurities was presented in [27]. Further analysis of deposits in real SPS can be useful for improving of SPS performances. Secondary Mass Spectroscopy (SIMS) can be important tool for such analysis.

Up to now it is not clearly how important CW RF discharge (used in the internal antenna source for pulsed discharge triggering) for cesium preservation (it can keep cesium in ionized state and prevent Cs+ ion escaping by extraction voltage).

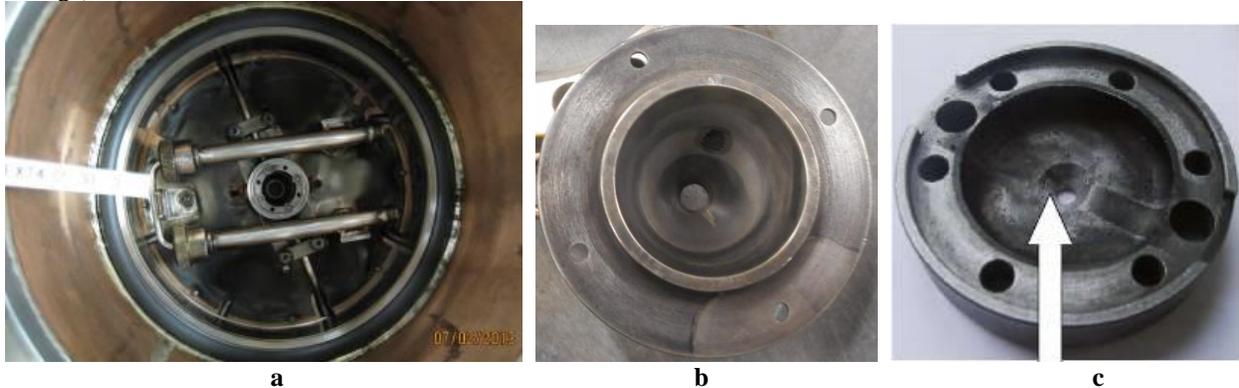

        a                                      b                                  c

**FIGURE 8**: a-Dark film deposition on the plasma plate components after 2 weeks of EA SPS stable operation: heat shield and converter cone (surfaces bombarded by intense fluxes of plasma are clean). **b-** Dark film deposition on the plasma plate components of SA SPS with high efficiency of H- ion production.
c-Deposition of dark film on the emitter cone surface of Penning discharge SPS.

Up to now it is not clearly how important this CW RF discharge for electrodes conditioning and for optimal carbon films deposition. During long time was not produced a long time operation of the external antenna SPS without degradation of H- beam current. As possible reasons were discussed harmful impurities emission from the AlN chamber and cesiated surface poisoning by these impurities or Cs desorbtion by heavy ions [3]. Other reason can be a chemical sputtering-formation of gas molecules $NH_3$ for hydrogen plasma interaction with AlN chamber. $NH_3$ have a larger dipole moment and can adsorbed on the emitter surface with increase of the WF and degrade H- emission.

Recently the stable (persistent) operation of the EA SPS was observed and need understand main factors of this transformation. Dark film deposition to the plasma plate components after 2 weeks of EA SPS stable operation: heat shield and converter cone (surfaces bombarded by intense fluxes of plasma are clean) shown in Fig. 8 a is very similar to deposits with internal antenna sources shown in Figs. 3-5.

Dark film is deposited to the molybdenum cone of converter but does not deposited to stainless steel aperture following the cone aperture.

The H- beam intensity was increased by increase of cone aperture without increase of SS aperture.

Dark film deposition was observed on surfaces of new design of converter cone in saddle Antenna RF SPS shown in Fig. 8 b. In this SA SPS was reached high plasma density on the axis at low RF power and high efficiency of negative and positive ion generation. In Fig. 8 c is shown dark film in CW Penning discharge converter cone from in BINP SBRAS [28].

In this case for the part of connection of a hydrogen bottle with ion source was used a plastic tube, which can be source of hydrocarbons. With a "clean" all metal gas system the converter cone was cleaned by dense plasma flax and cesium deposition was lost very fast with decrease of negative ion signal It is interesting to analyze the surface of emitters by SIMS.

It can be important a photo induced desorption and intercalation of alkali atoms on graphite observed in [22]. For K on graphite the desorption cross section has a resonance maximum $\sim 2 \cdot 10^{-20}$ cm$^2$ at photon energy 4.9 eV.

Further analysis of deposits in real SPS can be useful for improving of SPS performances. Secondary Mass Spectroscopy (SIMS) can be important tool for such analysis.

Optimization of deposited films can be used for improving of SPS performances.

Using of the converter/emitters made from balk graphite can be not practical because it has low thermal conductivity and can intercalate too much cesium in volume with low concentration on the surface.

The collector current is increase with increase of a magnetic field up to Um ~4 V, and decrease with further increase of magnetic field because a plasma flux is compressed to the emission aperture and interaction of plasma flux with a collar surface is decreases. The specific power efficiency of negative ion beam production in CW mode is up to Spe = 20 mA/cm$^2$ kW. (In the existing RF SPS the Spe ~ 4-6 mA/cm$^2$ kW; in the TRUIMF filament arc discharge negative ion source the best Spe is about 2 mA/cm$^2$ kW; in a compact Penning discharge SPS the Spe is 100 mA/cm$^2$ kW).

# REFERENCES

[1] V. Dudnikov, SU patent C1.H013/04, No. 411542 (10 March 1972). http://www.findpatent.ru/patent/41/411542.html

[2] Y. Belchenko and V. Dudnikov, "Surface negative ion production in ion sources," in *Production and Application of Light Negative Ions, 4th European Workshop*, edited by W. Graham (Belfast University, Belfast, 1991), pp. 47–66; Y. Belchenko, Rev. Sci. Instrum. **64**, 1385 (1993).

[3] V. Dudnikov, "Forty Years of Surface Plasma Sources Development", The Review of scientific instruments, V: 83 , 02A708, (2012).

[4] Y. I. Belchenko, G. I. Dimov, and V. G. Dudnikov, *Symposium on the Production and Neutralization of Negative Hydrogen Ions and Beams, Brookhaven, 1977* (Brookhaven National Laboratory (BNL), Upton, NY, 1977), pp. 79–96.

[5] Y. I. Belchenko, G. I. Dimov, and V. G. Dudnikov Nucl. Fusion **14**(1), 113 (1974).

[6] Martin P. Stockli, B. X. Han, T. W. Hardek, Y. W. Kang, S. N. Murray, T. R. Pennisi, C. Piller, M. Santana, and R.Welton, "Producing persistent, high-current, high-duty-factor H- beams for routine 1 MW operation of Spallation Neutron Source", Review of Scientific Instruments 83, 02A732 (2012).

[7] Martin P. Stockli, "Plasma-Wall Interactions in the Cesiated SNS H- Ion Source", Journal of Physics: Conference Series 399, 012001(2012).

[8] Stockli, M. P.; Han, B.; Murray, S. N., et al., "Ramping up the Spallation Neutron Source beam power with the H- source using 0 mg Cs/day", REVIEW OF SCIENTIFIC INSTRUMENTS, 81, 02A729 (2010

[9] V. Dudnikov, et al., "Carbon Film in RF Surface Plasma Source with cesiation", Report in ICIS 2017, Geneva (2017).

[10] E. V. Rut'kov and A. Ya. Tontegode, Pis'ma Zh. Tekh. Fiz. **7**, 1122 (1981) [Sov. Tech. Phys. Lett. **7**, 480 (1981)].

[11] A. Ya. Tontegode and E. V. Rut'kov, Usp. Fiz. Nauk **163**, 57 (1993) [Phys.-Usp. **36**, 1053 (1993)].

[12] N. R. Gall, E. V. Rut'kov, and A. Ya. Tontegode, Int. J. Mod. Phys. B **11**, 1865 (1997).

[13] A. Ya. Tontegode, Prog. Surf. Sci. **38**, 201 (1991).

[14] E. V. Rut'kov, A. Ya. Tontegode, and M. M. Usufov, Phys. Rev. Lett. **74**, 758 (1995).

[15] . Martin P. Stockli, Baoxi Han, Syd Murray, Terry Pennisi, Manuel Santana, Robert Welton, and Harry Meyer III, "Cs Puzzles with the SNS H- Source", *Cs'13 a Mini Workshop on Controlling Cs, Chiba, Japan September 14, (2013)*.

[16] B. Feuerbacher, B. Fitton, J. Appl. Phys. 43 (1972) 1563.

[17] in: D.R. Lide (Ed.), CRC Handbook of Chemistry and Surf. Sci. 279 (1992) 149. Physics (1995) CRC Press, Florida, 1996.

[18] L. Osterlund, D.V. Chakarov, B. Kasemo, Potassium adsorption on graphite (0001), Surface Science, V. 420, Issue 2, p. 174-189 (1999).

[19] V. Dudnikov., P. Chapovsky , A. Dudnikov, "Cesium Control and Diagnostics In Surface Plasma Negative Ion Sources", Rev. Sci. Instrum. 81, 02A714, 2010.

[20] M.A. Gleeson, A.W. Kleyn, Effects of Cs-adsorption on the scattering of low energy hydrogen ions from HOPG, Surface Science 420 (1999) 174–189.

[21] Sergei Yu. Davydov, "Alkali metal adsorption on graphite: Calculation of a work function variation in the Anderson–Newns–Muscat model", Applied Surface Science 257 (2010) 1506–1510.

[22] . M. Breitholz, J. Algdal, T. Kihlgren, S.A. Lindgren, L. Wallden, Phys. Rev. B 70, 125108(2004).

[23] R. Gutser, C. Wimmer, and U. Fantz, "Work function measurements during plasma exposition at conditions relevant in negative ion sources for the ITER neutral beam injection", Rev. Sci. Instrum. 82, 023506 (2011).

[24] V. Dudnikov et al., IPAC 2010, Kyoto, 2010, THPEC072 (2010).

[25] V. Dudnikov, R. P. Johnson, « Cesiation in highly efficient surface plasma sources », Physical Review Special Topics - Accelerators and Beams (Vol.14, No.5 (2011): http://link.aps.org/doi/10.1103/PhysRevSTAB.14.054801

[26] V.Z.Kaibyshev, V.A.Koryukin and V.P.Obrezumov, Atomnaya Energiya, **69**, No.3, 196 (1990).

[27] A. Ueno, H. Oguri, K. Ikegami, Y. Namekawa, and K. Ohkoshi, Rev. Sci. Instrum. **81**, 02A715 (2010).

[28] Yu. Belchenko, A. Gorbovsky, A. Sanin, and V. Savkin, "The 25 mA continuous-wave surface-plasma source of H ions", Review of Scientific Instruments 85, 02B108 (2014).